\documentclass[superscriptaddress,aps,pra,preprint]{revtex4-1}
\usepackage{graphicx}
\usepackage{bm}

\begin{document}

\title{Three-level Haldane-like model on dice optical lattice}

\date{\today}

\author{T.~Andrijauskas}
\author{E.~Anisimovas}
\author{M.~Ra\v{c}i\={u}nas}
\author{A.~Mekys}
\author{V.~Kudria\v{s}ov}
\affiliation{Institute of Theoretical Physics and Astronomy,
Vilnius University, A.~Go\v{s}tauto 12, Vilnius LT-01108, Lithuania}

\author{I.~B.~Spielman}
\affiliation{Joint Quantum Institute, University of Maryland, College Park,
Maryland 20742-4111, 20742, USA}
\affiliation{National Institute of Standards and Technology, Gaithersburg,
Maryland 20899, USA.}

\author{G.~Juzeli\={u}nas}
\affiliation{Institute of Theoretical Physics and Astronomy,
Vilnius University, A.~Go\v{s}tauto 12, Vilnius LT-01108, Lithuania}

\begin{abstract}
We consider ultracold atoms in a two-dimensional optical lattice of
the dice geometry in a tight-binding regime. The atoms experience
a laser-assisted tunneling between the nearest neighbor sites of
the dice lattice accompanied by the momentum recoil. This allows one
to engineer staggered synthetic magnetic fluxes over plaquettes, and
thus pave a way towards the realization of topologically nontrivial
band structures. In such a lattice the real-valued next-neighbor
transitions are not needed to reach a topological regime. Yet, such
transitions can increase a variety of the obtained topological phases.
The dice lattice represents a triangular Bravais lattice with a three-site
basis consisting of a hub site connected to two rim sites. As a consequence,
the dice lattice supports three energy bands. From this point
of view, our model can be interpreted as a generalization of the paradigmatic
Haldane model which is reproduced if one of the two rim sub-lattices
is eliminated. We demonstrate that the proposed upgrade of the Haldane
model creates a significant added value, including an easy access to
topological semimetal phases relying only on the nearest neighbor coupling,
as well as enhanced topological band structures featuring Chern numbers
higher than one leading to physics beyond the usual quantum Hall effect. The 
numerical investigation is supported and complemented
by an analytical scheme based on the study of singularities in the
Berry connection.
\end{abstract}



\maketitle

\section{Introduction}

Optical lattices have firmly established themselves as a modern and
versatile tool to study fundamental physics in a clean environment
with various physical parameters being under experimentalist's control
and often extensively tunable~\cite{Bloch2008a,Lewenstein2007,Windpassinger2013}.
One is typically interested in implementing a paradigmatic Hamiltonian
that clearly demonstrates a particular phenomenon or an effect. A
list of recent successes features, to mention just a few examples,
realization of the Harper-Hofstadter \cite{Aidelsburger2013,miyake2013,Hofstadter1976}
and Haldane models~\cite{Jotzu2014}, direct observation and control
of the Dirac points~\cite{Tarruell2012}, creation of artificial
magnetic fluxes via lattice shaking~\cite{Struck2012} and reproduction
of models of magnetism~\cite{Struck2013}, engineering of a spin-dependent optical lattice
resulting from a combination of Raman coupling and radio-frequency magnetic 
fields~\cite{Jimenez-Garcia2012}.

In particular, access to topological band structures is of enormous
interest~\cite{Hasan2010,Qi2011,Goldman2014RPP}. The presence of
the topological order is signaled by a non-zero Chern index reflecting
a non-vanishing integral of the Berry curvature over the entire two-dimensional
Brillouin zone. A topological band supported by a spatially periodic
optical lattice acts as a model of a Landau level. The unique band structure
consisting of a ladder of Landau levels defines an apparent insulator
with current-carrying edge states and has traditionally been associated
with the presence of an external magnetic field. In cold-atom setups,
however, the topological character becomes an intrinsic property of
the band and is not necessarily associated with the presence of a
physical magnetic field~\cite{Goldman2014RPP,Sheng2011}. Synthetic
fluxes piercing the lattice plaquettes may be imparted by the lattice
shaking~\cite{Struck2012,Goldman2014RPP,Eckardt:2005,Kolovsky2011,Goldman2014},
laser-assisted tunneling~\cite{Goldman2014RPP,Jaksch2003,Gerbier2010,Dalibard2011}
or using synthetic dimensions~\cite{Celi:2013}.

Many of the breakthroughs mentioned in the introductory paragraph
can be classified as mimicking or reproduction of phenomena known
from the condensed matter physics. However, significant contributions
from cold-atom systems to \emph{extending} the known physics should
also be recognized~\cite{Bloch2008a,Lewenstein2007,Goldman2014RPP,Dalibard2011,Buljan2014}. Perhaps the most obvious examples relate to the
construction of topological bands with the values of the Chern index
greater than one~\cite{Wang2011,Liu2012,Yang2012,Wang2012,Sterdyniak2013,Wang2011PRB,Sticlet2012},
which is a central topic of the present paper. The properties of such
a band is not a direct sum of the properties of several Landau levels,
and reach beyond the traditional (integer or fractional) quantum Hall
physics~\cite{Parameswaran2013,Bergholtz2013}.

Indeed, the study of bands with higher Chern numbers has been particularly
relevant in connection to the so-called fractional Chern insulators~\cite{Neupert2011,Regnault2011,Grushin2014}.
Although many-body interactions, which play the central role in these
studies, are beyond the scope of the present contribution, we stress
that many insights into the nature of the fractional topological states
were obtained from somewhat artificial lattice constructs often involving
many layers~\cite{Liu2012} or distant-neighbour hoppings~\cite{Yang2012,Wang2012,Sticlet2013}.
Ongoing efforts~\cite{Kol1993,Palmer2006,Moller2009,Moller2015} are also based on the
Harper-Hofstadter model that in principle supports subbands of arbitrarily high Chern
numbers. Here, one also has to defy rather stringent requirements posed by large magnetic
unit cells, low particle densities, and a large number of subbands implying small
topological band gaps~\cite{Moller2015}.
In the present paper we focus on exploring
the potential offered by relatively \emph{simple} and thus more realistic
lattice models. We construct a generalization of the Haldane model~\cite{Haldane1988,Alba2011,Juzeliunas2013Physics,Goldman2013,Anisimovas2014}
by coupling three rather than two triangular sub-lattices. In this
way, the honeycomb lattice featured in the Haldane model is upgraded
to the dice lattice~\cite{Sutherland1986,Bercioux2009,Moller2012,Rizzi2006,Burkov2006,Bercioux2011} which supports a three-band model with a clean
access to interesting topological configurations, such as bands characterized
by the Chern number equal to $2$. 
In the dice-lattice
model it is just a complex valued nearest-neighbor (NN) coupling that is sufficient
to generate a \emph{staggered} synthetic magnetic flux and
reach nontrivial setups including a topological semimetal
phase. On the other hand, for spatially periodic hexagonal lattices,
non-trivial phases can not be reached just by having the complex-valued
nearest-neighbor coupling, one should add a real-valued next-neighbor
coupling~\cite{Alba2011,Juzeliunas2013Physics,Goldman2013}.
Note that the dice lattice affected by a \emph{uniform} magnetic flux was 
used to demonstrate a novel and intriguing
mechanism of localization of wave packets in Aharonov-Bohm cages~\cite{Vidal1998,Abilio1999,Movilla2011}.

The paper is structured as follows. In Section~II, we introduce the
lattice geometry and derive the $3\times3$ momentum-space Hamiltonian
encapsulating the physics. Then, Section~III describes the obtained
results starting from phases obtained in the presence of NN couplings
alone and proceeding to more complex configurations requiring next-nearest
neighbor (NNN) transitions. We conclude with a brief summarizing
Section~V.

\section{The model}

\subsection{Lattice geometry\label{sub:Lattice-geometry}}

\begin{figure}
\begin{centering}
\includegraphics[width=0.45\textwidth]{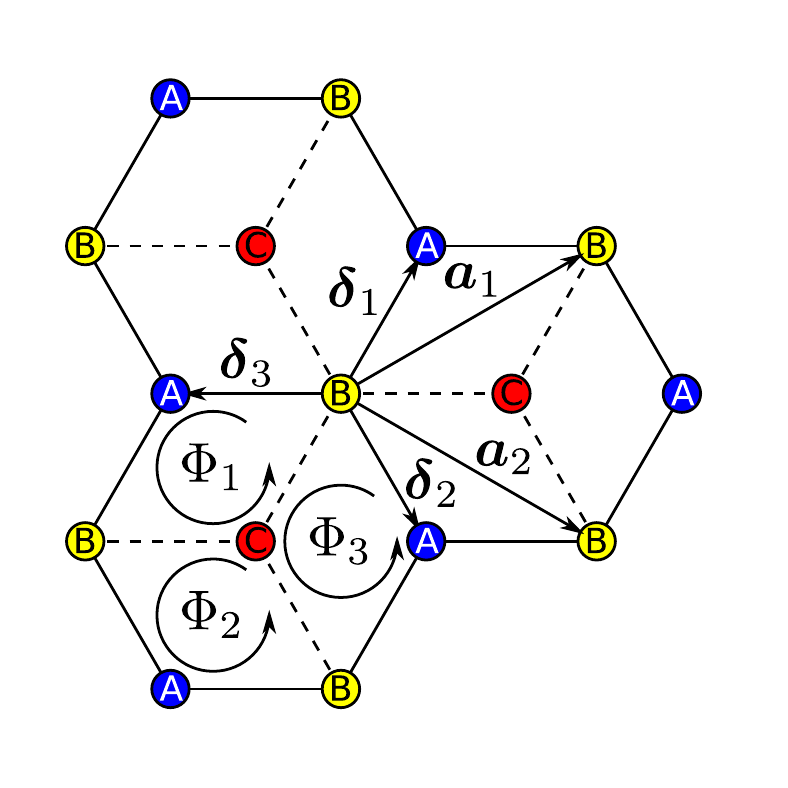}\hfill{}\includegraphics[width=0.45\textwidth]{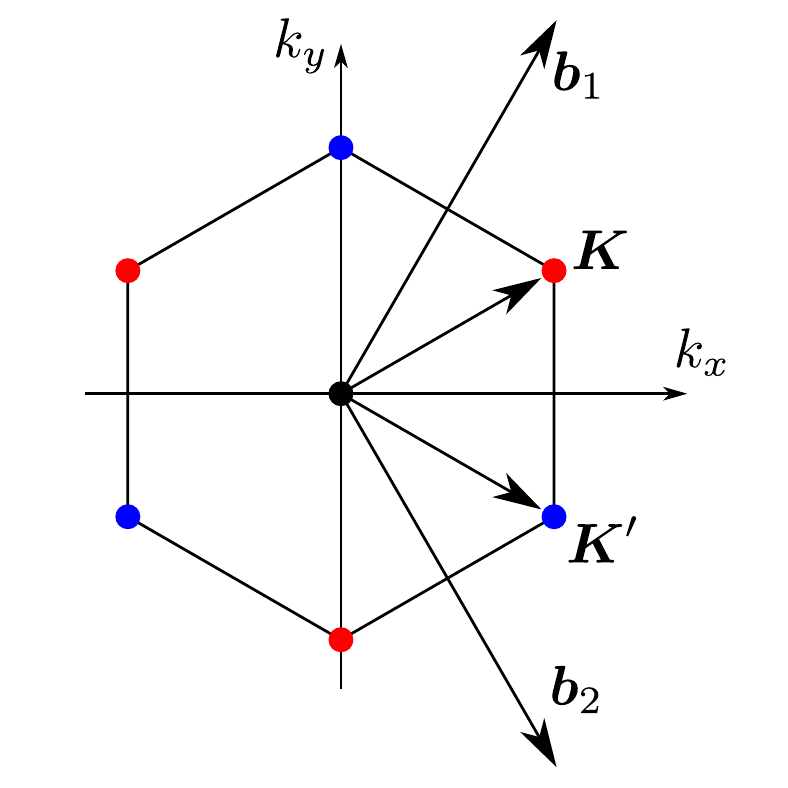}
\par\end{centering}

\protect\caption{(Color online) \emph{Left}: dice lattice. The blue, green and red sites correspond
to three different triangular sub-lattices A, B and C. Solid lines
show couplings between the sites A and B. Dashed lines show couplings
between the sites B and C. The primitive lattice vectors are $\bm{a}_{1}$
and $\bm{a}_{2}$. Nearest sites are connected with the vectors $\bm{\delta}_{1}$,
$\bm{\delta}_{2}$ and $\bm{\delta}_{3}$. \emph{Right}: hexagonal
first Brillouin zone of the reciprocal lattice defined by the primitive
reciprocal lattice vectors $\bm{b}_{1}$ and $\bm{b}_{2}$. Two inequivalent
corners are at the points $\bm{K}$ (red) and $\bm{K}^{\prime}$ (blue).}

\label{fig:lattice}
\end{figure}
We consider a dice lattice, which consists of three triangular sub-lattices.
One of them is called a \emph{hub} sub-lattice. It is coupled to
other two \emph{rim} sub-lattices, that in turn are not coupled with
each other. Let us denote the hub sub-lattice by B and the rim sub-lattice
by A and C. The vectors that connect the nearest lattice sites are
(Fig.~\ref{fig:lattice}):
\begin{equation}
\bm{\delta}_{1}=\frac{a}{2}(\bm{e}_{x}+\sqrt{3}\bm{e}_{y}),\quad\bm{\delta}_{2}=\frac{a}{2}(\bm{e}_{x}-\sqrt{3}\bm{e}_{y}),\quad\bm{\delta}_{3}=-a\bm{e}_{x}\,,\label{eq:nn-vectors}
\end{equation}
where $a$ is the distance between two such sites. The elementary lattice
vectors
\begin{equation}
\bm{a}_{1}=a(3\bm{e}_{x}+\sqrt{3}\bm{e}_{y})/2,\quad\bm{a}_{2}=a(3\bm{e}_{x}-\sqrt{3}\bm{e}_{y})/2\label{eq:lattice-vectors}
\end{equation}
define a rhombic elementary cell. The set
of lattice vectors $\bm{r}_{\bm{n}}=n_{1}\bm{a}_{1}+n_{2}\bm{a}_{2}$
(with integers $n_{1}$ and $n_{2}$) span the hub sub-lattice
B (Bravais lattice). The two rim sub-lattices are defined in the following
way. The first rim sub-lattice A is shifted from the hub sub-lattice
B by the vector $\bm{\delta}_{1}$ in such way that sub-lattices A
and B alone make a honeycomb lattice. The second rim sub-lattice C
is shifted to the opposite direction by $-\bm{\delta}_{1}$ (see Fig.~\ref{fig:lattice}).
Let us introduce a set of vectors, that span all the lattice sites:
\begin{equation}
\bm{r}_{\bm{n},s}=\bm{r}_{\bm{n}}+s\bm{\delta}_{1}.\label{eq:lattice-sites}
\end{equation}
Here the index $s=0,\pm1$ labels the three sub-lattices. The sites
of the hub sub-lattice ($s=0$) coincide with the lattice vectors:
$\bm{r}_{\bm{n},0}=\bm{r}_{\bm{n}}$. The sites of the rim sub-lattices
A and C shifted by $\pm\bm{\delta}_{1}$, i.e. $\bm{r}_{\bm{n},+1}=\bm{r}_{\bm{n}}+\bm{\delta}_{1}$
and $\bm{r}_{\bm{n},-1}=\bm{r}_{\bm{n}}-\bm{\delta}_{1}$.

It is convenient to introduce an additional lattice vector $\bm{a}_{3}=\bm{a}_{1}-\bm{a}_{2}$.
The set of the three lattice vectors $\bm{a}_{i}$ ($i=1,2,3$) together
with the opposite vectors $-\bm{a}_{i}$ connect all next-nearest
lattice sites, and can be related to $\bm{\delta}_{i}$ as: $\bm{a}_{1}=\bm{\delta}_{1}-\bm{\delta}_{3}$,
$\bm{a}_{2}=\bm{\delta}_{2}-\bm{\delta}_{3}$ and $\bm{a}_{3}=\bm{\delta}_{1}-\bm{\delta}_{2}$.

The basic reciprocal lattice vectors
\begin{equation}
\bm{b}_{1}=\frac{2\pi}{3a}(\bm{e}_{x}+\sqrt{3}\bm{e}_{y}),\quad\bm{b}_{2}=\frac{2\pi}{3a}(\bm{e}_{x}-\sqrt{3}\bm{e}_{y})\label{eq:recip-lattice-vectors}
\end{equation}
are orthogonal to the lattice vectors, $\bm{a}_{i}\cdot\bm{a}_{j}=2\pi\delta_{ij}$,
$i,j=1,2$. The first Brillouin zone is hexagonal with two inequivalent
corners $\bm{K}$ and $\bm{K}^{\prime}$ positioned at $\bm{K}=(2\bm{b}_{1}+\bm{b}_{2})/3$
and $\bm{K}^{\prime}=(\bm{b}_{1}+2\bm{b}_{2})/3$. In terms of the
Cartesian coordinates these points are given by 
\begin{equation}
\bm{K}=\frac{2\pi}{9a}(3\bm{e}_{x}+\sqrt{3}\bm{e}_{y}),\quad\bm{K}^{\prime}=\frac{2\pi}{9a}(3\bm{e}_{x}-\sqrt{3}\bm{e}_{y}),\label{eq:BZ-corners}
\end{equation}
as one can see in Fig.~\ref{fig:lattice}.

\subsection{Tight-binding model}

We shall make use of the tight-binding model in which the single particle
states $|\bm{r}_{\bm{n},s}\rangle$ represent the Wannier-type wave-functions
localized at each lattice site $\bm{r}_{\bm{n},s}$, with $s=0,\pm1$
being the sub-lattice index. In the language of the second quantization
these single-particle states read $|\bm{r}_{\bm{n},s}\rangle=c^{\dag}(\bm{r}_{\bm{n},s})|{\rm vac}\rangle$,
where $|{\rm vac}\rangle$ is the Fock vacuum state, $c^{\dag}(\bm{r}_{\bm{n},s})$
and $c(\bm{r}_{\bm{n},s})$ being the creation and annihilation operators
of an atom in the corresponding localized state. 

The full Hamiltonian of the system consist of three terms,
\begin{equation}
H=H_{1}+H_{2}+H_{3}.\label{eq:full-Hamiltonian}
\end{equation}
The first term $H_{1}$ describes the laser-assisted tunneling  
\cite{Goldman2014RPP,Goldman2014,Jaksch2003,Gerbier2010,Dalibard2011,Alba2011,Goldman2013,Ruostekoski:2002} 
of atoms between the sites of the hub sub lattice B ($s=0$) and its nearest
neighboring sites that belong to the rim sub-lattices A and C with
$s=\pm1$:
\begin{equation}
H_{1}=\sum_{\bm{n}}\sum_{s=\pm1}J^{(s)}\sum_{i=1}^{3}{\rm e}^{{\rm i}\bm{p}_{s}\cdot(\bm{r}_{\bm{n}}+s\bm{\delta}_{i}/2)}c^{\dag}(\bm{r}_{\bm{n}})c(\bm{r}_{\bm{n}}+s\bm{\delta}_{i})+{\rm H.\, c.}\,,\label{eq:H-1}
\end{equation}
where $J^{(s)}$ are the coupling amplitudes.
Such generalization of dice optical lattice with two different hopping parameters
$J^{(+)}$ and $J^{(-)}$ is already discussed in \cite{Raoux2014}.
The laser assisted tunneling
is accompanied by the transfer of the recoil momentum $\bm{p}_{s}$
with $s=\pm1$, to be labelled simply by $\bm{p}_{\pm}\equiv\bm{p}_{\pm1}$.
In the present situation $\bm{p}_{+}$ can generally differ from $\bm{p}_{-}$
because the transitions between the different sub-lattices can be induced
by different lasers. Note that the nearest neighbor hopping alone
is sufficient to generate fluxes through rhombic plaquettes 
\begin{equation}
\Phi_{i}=\pm(\bm{p}_{+}-\bm{p}_{-})\cdot\bm{a}_{i}/2\,,\label{eq:Phi_i}
\end{equation}
 with $\bm{a}_{i}$ representing a diagonal vector of the plaquette
in question. Yet the magnetic flux over the whole hexagonal plaquette
remains zero.

The second term $H_{2}$ takes into account the tunneling between
the next-nearest neighboring sites belonging to the same sub-lattice
with $s=0,\pm1$:
\begin{equation}
H_{2}=\sum_{\bm{n}}\sum_{s=0,\pm1}J_{2}^{(s)}\sum_{i=1}^{3}c^{\dag}(\bm{r}_{\bm{n},s})c(\bm{r}_{\bm{n},s}+\bm{a}_{i})+{\rm H.\, c.}\label{eq:H-2}
\end{equation}
This term describes the usual (not laser-assisted) hopping transitions
between nearest sites in each of the three triangular sub-lattices,
and $J_{2}^{(s)}$ with $s=0,\pm1$ are the corresponding matrix elements
for the tunneling between the atoms belonging to the $s$-th sub-lattice.

The third term $H_{3}$ describes the energy mismatch for the particles
populating the different sub-lattices:
\begin{equation}
H_{3}=\sum_{\bm{n}}\sum_{s=0,\pm1}\varepsilon_{s}c^{\dag}(\bm{r}_{\bm{n},s})c(\bm{r}_{\bm{n},s}).\label{eq:H-3}
\end{equation}
The on-site energies $\varepsilon_{s}$ are the diagonal matrix elements
of the Hamiltonian in the basis of the Wannier states. Without a loss
of generality we can take the on-site energy of the hub sub-lattice
B to be zero: $\varepsilon_{0}=0$. The on-site energies of other
\emph{rim} sub-lattices are to be labeled as $\varepsilon_{\pm1}\equiv\varepsilon_{\pm}$. 

Since the first term $H_{1}$ involves complex phase factors that
depend on the elementary cell number $\bm{n}$, the full Hamiltonian
$H$ is not translationally invariant. Yet, we will transform the annihilation
operators according to $c(\bm{r}_{\bm{n},0})\to c(\bm{r}_{\bm{n},0})$
and $c(\bm{r}_{\bm{n},s})\to c(\bm{r}_{\bm{n},s})\exp(-{\rm i}\bm{p}_{s}\cdot\bm{r}_{\bm{n}})$
with $s=\pm1$, and perform the corresponding transformation for the
creation operators. This gauge transformation makes the full Hamiltonian
(\ref{eq:full-Hamiltonian}) translationally invariant.

Transition to the reciprocal space is carried out by introducing new
operators
\begin{equation}
c_{s}(\bm{k})=\frac{1}{\sqrt{N}}\sum_{\bm{k}}c(\bm{r}_{\bm{n},s}){\rm e}^{-{\rm i}\bm{k}\cdot\bm{r}_{\bm{n}}}\,,\quad c(\bm{r}_{\bm{n},s})=\frac{1}{\sqrt{N}}\sum_{\bm{k}}c_{s}(\bm{k}){\rm e}^{{\rm i}\bm{k}\cdot\bm{r}_{\bm{n}}}\,,\label{eq:Transformation-momentum-space}
\end{equation}
together with the Hermitian conjugated creation operators $c_{s}^{\dagger}(\bm{k})$.
Here $N$ is a number of elementary cells in the quantisation area,
and the vectors $\bm{r}_{\bm{n}}=\bm{r}_{\bm{n},0}$ (defined in the
Subsec.~\ref{sub:Lattice-geometry}) are located at the sites of
the hub lattice. In terms of the new operators the Hamiltonian (\ref{eq:full-Hamiltonian})
splits into its $\bm{k}$-components: 
\begin{equation}
H=\sum_{\bm{k}}H(\bm{k})\,,\quad H(\bm{k})=\left[\begin{array}{ccc}
c_{+}^{\dag}(\bm{k}) & c_{0}^{\dag}(\bm{k}) & c_{-}^{\dag}(\bm{k})\end{array}\right]\mathcal{H}(\bm{k})\left[\begin{array}{c}
c_{+}(\bm{k})\\
c_{0}(\bm{k})\\
c_{-}(\bm{k})
\end{array}\right],\label{eq:H(k)}
\end{equation}
where $\mathcal{H}(\bm{k})$ is a $3\times3$ matrix: 
\begin{equation}
\mathcal{H}(\bm{k})=\left[\begin{array}{ccc}
\varepsilon_{+}+2J_{2}^{(+)}f(\bm{k}-\bm{p}_{+}) & J^{(+)}g(\bm{k}-\bm{p}_{+}/2) & 0\\
J^{(+)}g^{*}(\bm{k}-\bm{p}_{+}/2) & 2J_{2}^{(0)}f(\bm{k}) & J^{(-)}g(\bm{k}-\bm{p}_{-}/2)\\
0 & J^{(-)}g^{*}(\bm{k}-\bm{p}_{-}/2) & \varepsilon_{-}+2J_{2}^{(-)}f(\bm{k}-\bm{p}_{-})
\end{array}\right]\,.\label{eq:H-matrix}
\end{equation}

Here we also added an extra phase factor to the transformed operators
$c_{s}(\bm{k})\to c_{s}(\bm{k}){\rm e}^{{\rm i}\bm{p}_{s}\cdot s\bm{\delta}_{1}/2}$.
The functions 
\begin{equation}
f(\bm{k})=\sum_{i=1}^{3}\cos(\bm{k}\cdot\bm{a}_{i}),\quad g(\bm{k})={\rm e}^{{\rm i}\bm{k}\cdot\bm{\delta}_{1}}\sum_{i=1}^{3}{\rm e}^{-{\rm i}\bm{k}\cdot\bm{\delta}_{i}}\label{eq:f-and-g-functions}
\end{equation}
entering Eq.~(\ref{eq:H-matrix}) are translationally symmetric in
the reciprocal space 
\begin{equation}
f(\bm{k}+\bm{G})=f(\bm{k})\,,\quad g(\bm{k}+\bm{G})=g(\bm{k})\,,\label{eq:f-and-g-translation}
\end{equation}
where $\bm{G}=n_{1}\bm{b}_{1}+n_{2}\bm{b}_{2}$ is a reciprocal lattice
vector, $n_{1}$ and $n_{2}$ being integers.  Consequently the matrix-Hamiltonian
$\mathcal{H}(\bm{k})$ is also fully translationally invariant in the
reciprocal space $\mathcal{H}(\bm{k})=\mathcal{H}(\bm{k}+\mathbf{G})$.
Note that Berry curvature in general depends on the choice of Fourier transformation
(\ref{eq:Transformation-momentum-space}), while the corresponding Chern number does not
\cite{Jackson2014,Dobardzic2015}.
Furthermore, the functions $f(\bm{k})$ and $g(\bm{k})$ obey the
following reflection properties 
\begin{equation}
f(\bm{k})=f(-\bm{k})\,,\quad g(\bm{k})=g^{*}(-\bm{k})\,.\label{eq:f-and-g-reflection}
\end{equation}
All this helps to consider various symmetries of the matrix-Hamiltonian
(\ref{eq:H-matrix}).

\section{Phases of non-interacting fermions}

\subsection{Chern numbers and symmetries of the system}

Since the momentum-space Hamiltonian (\ref{eq:H-matrix}) represents
a three level system, there are three energy bands characterized by energies $E_{n}(\bm{k})$,
with $n=1,2,3$. Each energy band has a Chern number $c_{n}$
to be defined in Eq.~(\ref{eq:Chern-number}).
We also identify two possible band gaps. The first
band gap $\Delta_{12}$ measures the energy between the first ($n=1$) and
second ($n=2$) bands, the second band gap $\Delta_{23}$ corresponding to the energy
between the second ($n=2$) and the third ($n=3$) bands.

The Chern number $c_{n}$ for the $n$-th band is defined in terms
of a surface integral of a Berry curvature over the first Brillouin
zone (FBZ)~\cite{Goldman2014RPP,Xiao2010}:
\begin{equation}
c_{n}=-\frac{1}{2\pi}\int_{{\rm FBZ}}{\rm d}^{2}k\, F_{n}(\bm{k}).\label{eq:Chern-number}
\end{equation}
The Berry curvature $F_{n}(\bm{k})$ can be expressed in terms of
the eigenvectors $|u_{n,\bm{k}}\rangle$ of the reciprocal space Hamiltonian (\ref{eq:H-matrix})
as 
\begin{equation}
F_{n}(\bm{k})={\rm i}\left(\frac{\partial}{\partial k_{x}}\langle u_{n,\bm{k}}|\right)\left(\frac{\partial}{\partial k_{y}}|u_{n,\bm{k}}\rangle\right)-{\rm i}\left(\frac{\partial}{\partial k_{y}}\langle u_{n,\bm{k}}|\right)\left(\frac{\partial}{\partial k_{x}}|u_{n,\bm{k}}\rangle\right)\,.\label{eq:Berry-curvature}
\end{equation}
It is well defined as long the eigen-energies $E_{n}(\bm{k})$ are
not degenerate for any fixed value of $\bm{k}$. Therefore the Chern
number $c_{n}$ can be ascribed to the $n$-th band if the latter
does not touch any other bands. If the Fermi energy is situated in
a band gap, the Chern number is directly related to Hall conductivity
due to chiral edge states of the occupied bands \cite{Hatsugai1993}
via $\sigma_{xy}=-e^{2}c_{n}/\hbar$ \cite{Kohmoto1985,Kohmoto1989,Thouless1982}.
For numerical calculation we make use of the discretized version of
the Berry curvature (\ref{eq:Berry-curvature}) described in \cite{Fukui2005}.

For both \emph{rim} sub-lattices $A$ and $C$, we set on-site energies
of to be symmetrically shifted away from the zero point $\varepsilon_{+}=-\varepsilon_{-}=\varepsilon$.
We also take the tunneling amplitudes to be equal $J^{(+)}=J^{(-)}=J$,
$J_{2}^{(+)}=J_{2}^{(0)}=J_{2}^{(-)}=J_{2}$ and assume the recoil
momenta to be opposite $\bm{p}_{+}=-\bm{p}_{-}=\bm{p}$ for both \emph{rim}
sub-lattices $A$ and $C$. The choice of opposite recoil momenta ensures
the maximum flux, because the magnetic flux through a rhombic plaquette
$\Phi_{i}$ given by Eq.~(\ref{eq:Phi_i}) is proportional to the
difference of these vectors. Under these conditions, the matrix representation
of the $\bm{k}$-space Hamiltonian becomes
\begin{equation}
\mathcal{H}(\bm{k})=\left[\begin{array}{ccc}
\varepsilon+2J_{2}f(\bm{k}-\bm{p}) & Jg(\bm{k}-\bm{p}/2) & 0\\
Jg^{*}(\bm{k}-\bm{p}/2) & 2J_{2}f(\bm{k}) & Jg(\bm{k}+\bm{p}/2)\\
0 & Jg^{*}(\bm{k}+\bm{p}/2) & -\varepsilon+2J_{2}f(\bm{k}+\bm{p})
\end{array}\right].\label{eq:H-matrix-simplified}
\end{equation}
This form of the Hamiltonian exhibits some symmetries. The first symmetry
involves inversion of the on-site energies $\varepsilon\to-\varepsilon$
followed by the unitary transformation that changes the first row with
the third one (i.e. interchanges the \emph{rim} sub-lattices $A$ and
$C$), as well as the momentum inversion $\bm{k}\to-\bm{k}$. Using
the reflection properties of the functions $f$ and $g$ given by
Eq. (\ref{eq:f-and-g-reflection}), one arrives at the same Hamiltonian
(\ref{eq:H-matrix-simplified}). The second symmetry is $J\to-J$,
which is a simple gauge transformation. Using these two symmetries
we see that the change $J_{2}\to-J_{2}$ gives $\mathcal{H}(\bm{k})\to-\mathcal{H}(\bm{k})$.
To sum up, all the three mentioned symmetries are: $(\varepsilon\to-\varepsilon,\mathcal{H}\to\mathcal{H})$,
$(J\to-J,\mathcal{H}\to\mathcal{H})$ and $(J_{2}\to-J_{2},\mathcal{H}\to-\mathcal{H})$.

\subsection{Numerical analysis}

In this Subsection, we numerically study the Chern phases of non-interacting
fermions. In order to present dependence of the Chern number on the
parameters $\varepsilon$, $J$, $J_{2}$ and $\bm{p}$ we adopt a
similar presentation of the phase diagram scheme as in \cite{Goldman2013}.
We choose the energy unit to be the nearest-neighbor tunneling amplitude
$J$. For the recoil momentum $\bm{p}$, we express the $p_{x}$ component
in the units of $K_{x}$ and the component $p_{y}$ in the units of
$K_{y}$, where $\bm{K}$ is one of the FBZ corners, defined in (\ref{eq:BZ-corners}).
In all the phase diagrams we present the dependence of the Chern number
$c_{n}=c_{n}(p_{x},p_{y})$ on the components of the recoil momentum
$\bm{p}$ using different colors for each possible values of $c_{n}$.
The areas corresponding to a topologically trivial phase with a zero
Chern number are shown in green ($c_{n}=0$). On the other hand, the
areas corresponding to non-trivial Chern phases are shown in yellow
($c_{n}=1$), red ($c_{n}=2$), cyan ($c_{n}=-1$) and blue ($c_{n}=-2$).
Additionally we display Chern number labels in all the presented phase diagrams.

\begin{figure}
\includegraphics[width=0.45\textwidth]{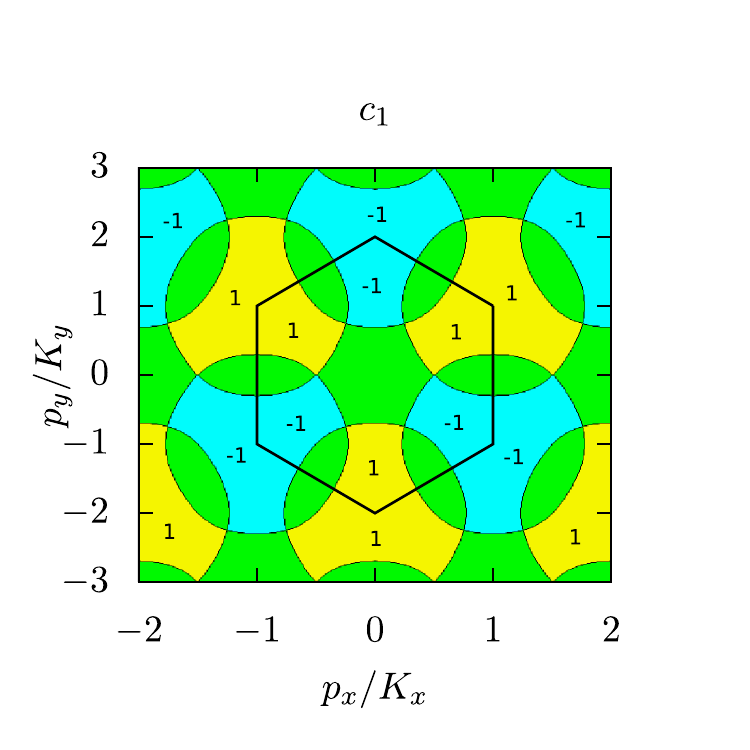}\hfill{}\includegraphics[width=0.45\textwidth]{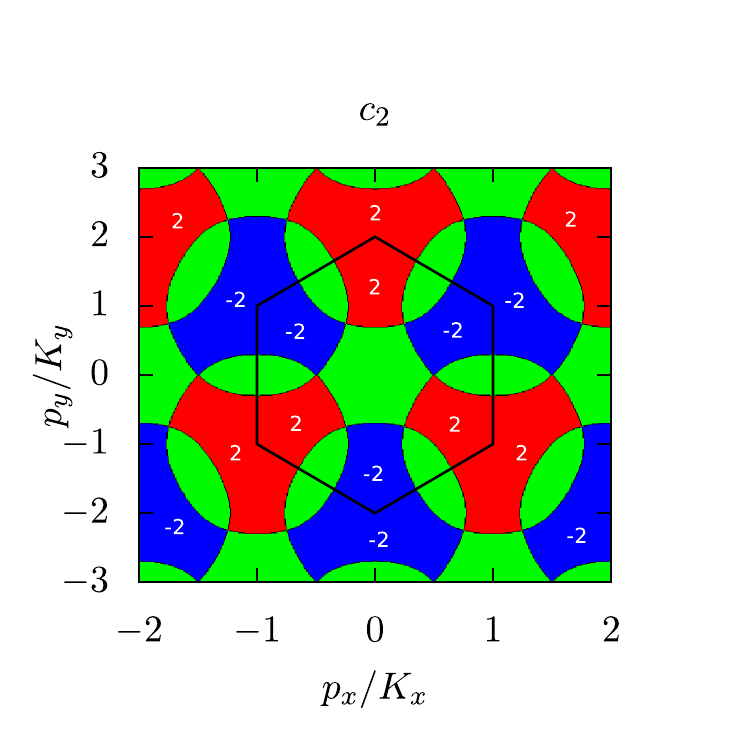}

\protect\caption{(Color online) Chern number dependence on the recoil momentum $\bm{p}$ in the case
$\varepsilon=J$ and $J_{2}=0$. In the left panel we present the
phase diagram of the lowest band Chern number $c_{1}$. In the right
panel we show corresponding phase diagram for the middle band. Since
the sum of Chern numbers over all three bands is zero, the third band
gives the same phase diagram as the first one ($c_{1}=c_{3}$). The
green regions
correspond to the Chern number zero. The yellow, red,
cyan and blue regions correspond to the Chern numbers 1, 2, -1 and
-2 respectively.
Nonzero Chern numbers are also displayed as labels.
The hexagon represents the FBZ in the $\bm{p}$-plane.}

\label{fig:semi-metal-121}
\end{figure}

First we characterize topological properties of the band structure
if there is no next-nearest neighbor coupling ($J_{2}=0$). In the
Fig.~\ref{fig:semi-metal-121} we show the Chern number phase diagrams
for $\varepsilon=J$. One can identify regions where the Chern numbers
are $\{c_{1},c_{2},c_{3}\}=\{0,0,0\}$, $\{-1,2,-1\}$ and $\{1,-2,1\}$.
In the first type of the regions (green color) we have topologically
trivial regions. In other regions there are non-zero Chern numbers with
band gaps $\Delta_{12}=\Delta_{23}=0$.
Analysis of the band structure
in these regions shows that the bands do not overlap
and touch indirectly.
Thus by filling the first one or the first two bands we
arrive at semi-metallic phase with non-zero Hall conductivity.
The typical spectrum of such non-trivial semi-metallic case is presented in
Fig.~\ref{fig:spectrum-121-202}.
The size of the non-trivial regions in the $\bm{p}$-plane depends on
the mismatch $\varepsilon$ of the on-site energies of A and C sublattices.
By increasing $\varepsilon$ from zero these regions immediately appear
around the points $\bm{p}=\bm{K}$
and become larger in size. For about $\varepsilon \approx J$ these regions have the largest area
as presented in the Fig.~\ref{fig:semi-metal-121} for $\varepsilon = J$.
For even larger values of $\varepsilon$ the non-trivial regions shrink back to the
points $\bm{K}$ and finally we
are left only with the trivial phase $\{ 0,0,0 \}$ everywhere.
The analytical treatment, presented in the section~IV gives the value
of $\varepsilon = \frac{3 \sqrt{2}}{2} J$
for which the semi-metal regions completely disappear.
For $J_2 = 0$ there are no other types of phases than
the trivial and semi-metallic discussed above.
Nonzero band gaps appear only in the regions of
trivial phase.

For the case $J_{2}=0$, the change $\bm{p}\to\bm{p}+\bm{G}$,
where $\bm{G}$ is the reciprocal lattice vector, corresponds to a
gauge transformation. Thus there is a symmetry $(\bm{p}\to\bm{p}+\bm{G},\mathcal{H}\to\mathcal{H})$.
In the phase diagram (Fig.~\ref{fig:semi-metal-121}) we also show
the FBZ in the $\bm{p}$-plane, which is a hexagon with two inequivalent
corners positioned at the points $\bm{K}$ and $\bm{K}^{\prime}$.

\begin{figure}
\includegraphics[width=0.45\textwidth]{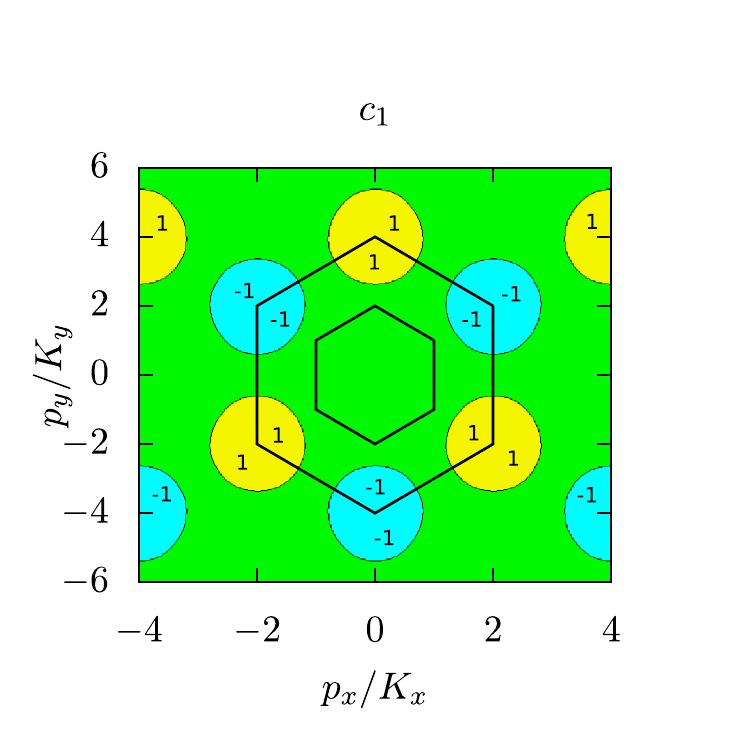}\hfill{}\includegraphics[width=0.45\textwidth]{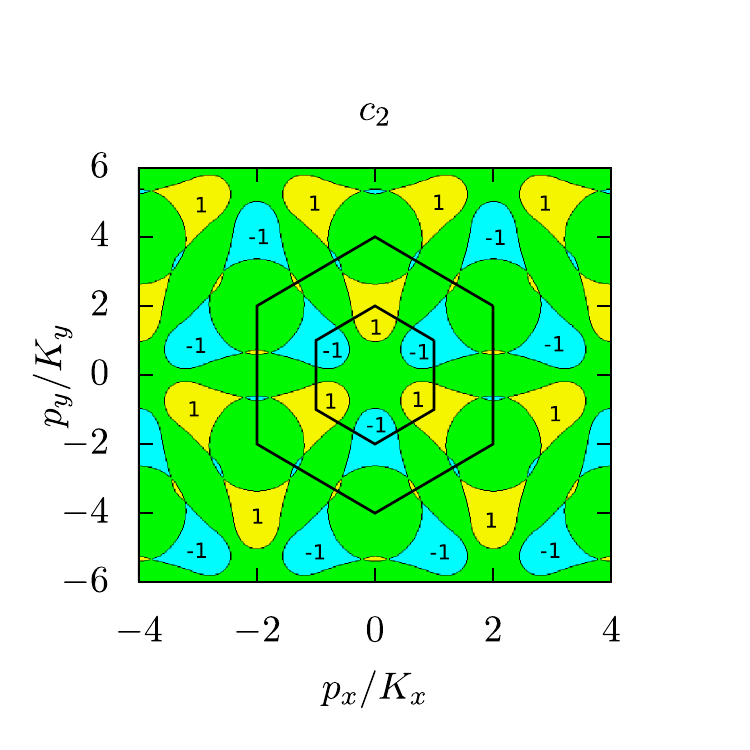}

\protect\caption{(Color online) Chern number dependence on the recoil momentum $\bm{p}$ in the case
$\varepsilon=2J$ and $J_{2}=0.3J$. In the left panel we present
the Chern number $c_{1}$ of the lowest band, while in the right panel
we show the Chern number $c_{2}$ of the middle band. For the third
band (not shown here) we have $c_{3}=-(c_{1}+c_{2})$. The green,
yellow, red, cyan and blue regions correspond to the Chern numbers
0, 1, 2, -1 and -2 respectively.
Nonzero Chern numbers are also displayed as labels.
A smaller hexagon shows the
FBZ corresponding to the case $J_{2}=0$. Since the introduction of non-zero
$J_{2}$ changes the periodicity of the $\bm{p}$-dependence, we also
show a bigger hexagon, which is now the FBZ in the $\bm{p}$-plane.}

\label{fig:insulator-101}
\end{figure}

\begin{figure}
\includegraphics[width=0.45\textwidth]{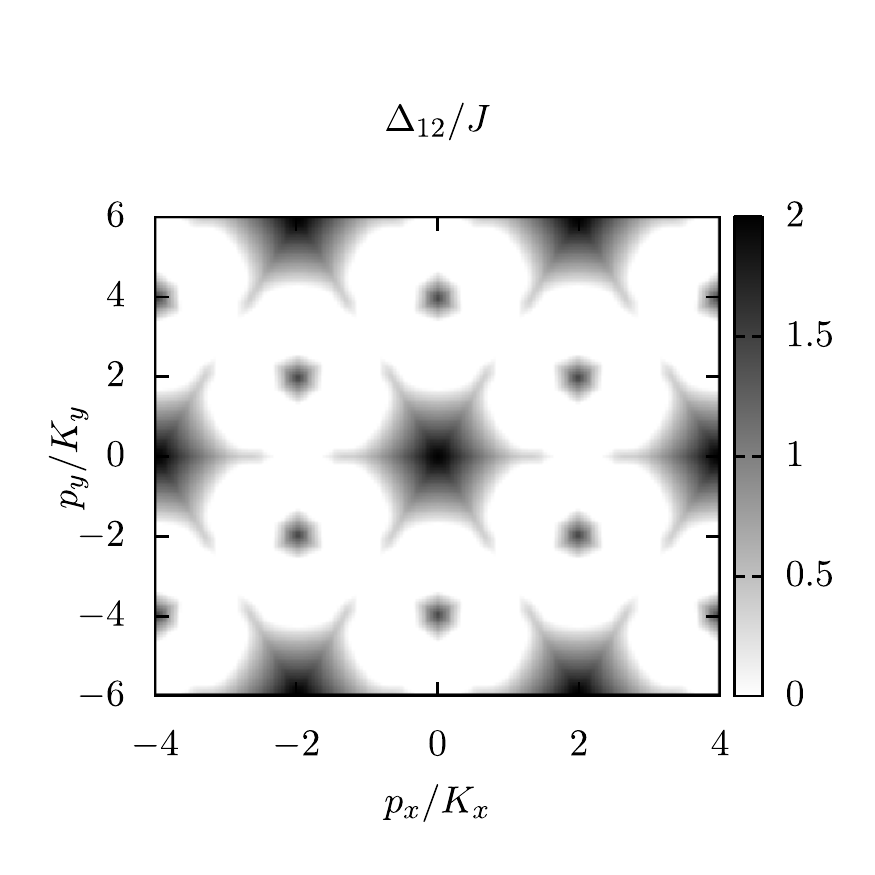}\hfill{}\includegraphics[width=0.45\textwidth]{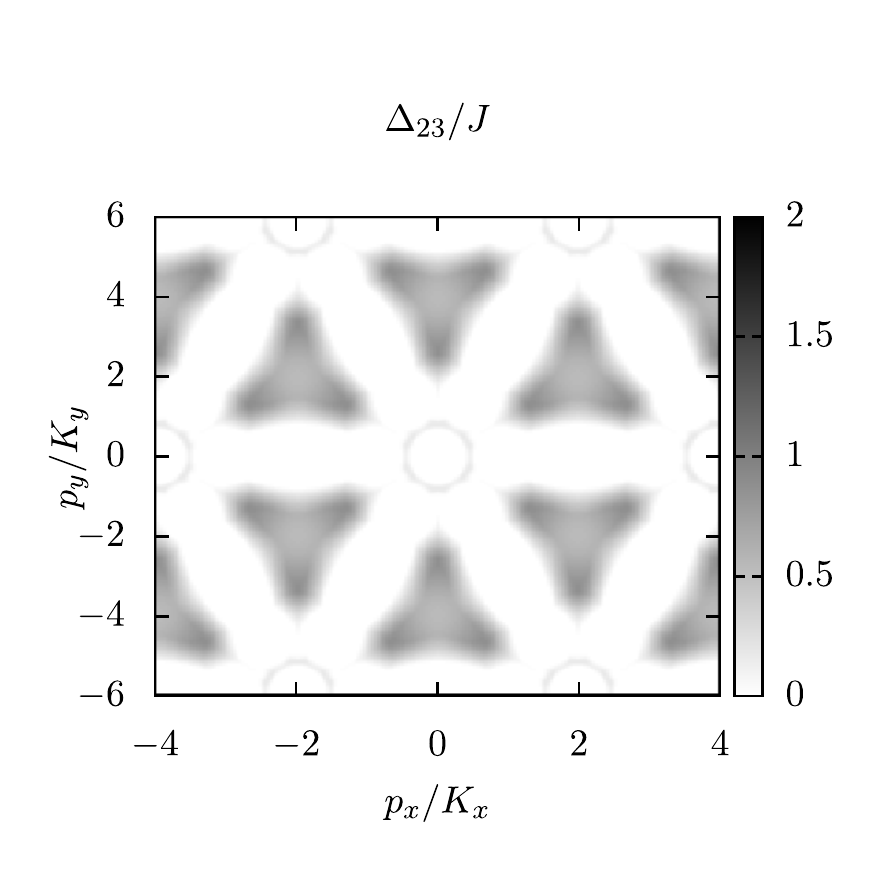}

\protect\caption{Dependence of the band gap on the recoil momentum $\bm{p}$ in the case
where $\varepsilon=2J$ and $J_{2}=0.3J$. In the left panel we present
the band gap $\Delta_{12}$ between the first and second bands. In the right panel
we show the band gap $\Delta_{23}$ between the second and third bands.}

\label{fig:gaps-101}
\end{figure}

Now let us analyze effects of the non-zero next-nearest neighbor
coupling. For this we set $J_{2}=0.3J$ and $\varepsilon=2J$. The
phase diagrams of the Chern numbers are presented in the Fig.~\ref{fig:insulator-101}.
We can see regions with the Chern numbers corresponding to trivial
phases $\{ 0,0,0 \}$ and phases $\{0,\pm1,\mp1\}$ and $\{\pm1,0,\mp1\}$.
In the latter two types of regions
we can find points corresponding to non-zero band gaps $\Delta_{12}>0$
and/or $\Delta_{23}>0$ (Fig.~\ref{fig:gaps-101}). This shows that there exist topological Chern
insulating phases.
For example at the point $\bm{p}=\bm{K}$, we have
the Chern numbers $\{0,-1,1\}$, the band gap between the
middle and highest bands being $\Delta_{23}\approx0.26J$.
Band widths in this case are about $3J$.
By positioning
the Fermi energy in the gap between the second and third bands  one arrives
at the Chern insulating phase. Another interesting point is $\bm{p}=2\bm{K}$,
which gives the
Chern numbers $\{-1,0,1\}$, the band gaps $\Delta_{12}\approx1.55J$
and $\Delta_{23}\approx0.54J$ and the band widths of about $2J$.
The bottom and top bands have non-zero
Chern numbers, while it is zero for the middle band. Depending on
the filling there are two types of topologically non-trivial phases.
If the Fermi energy is positioned in one of the band gaps, we get
a topological insulating phase. If the Fermi energy is situated within a band,
the band is partially filled and supports the Chern metal phase.
The discussed types of Chern number distributions over the bands are typical
when $J_2$ is non-zero and smaller than $J$ and $\varepsilon$.

In the case of non-zero NNN coupling
$J_{2}$ the translation
symmetry in the recoil momentum $\bm{p}$ is smaller than in the case
of zero NNN couplings: one has to shift the momentum by $2\bm{G}$
rather than $\bm{G}$. In the phase diagram presented in the Fig.~\ref{fig:insulator-101}
we show this by extending the FBZ, which is now a bigger hexagon.

\begin{figure}
\includegraphics[width=0.45\textwidth]{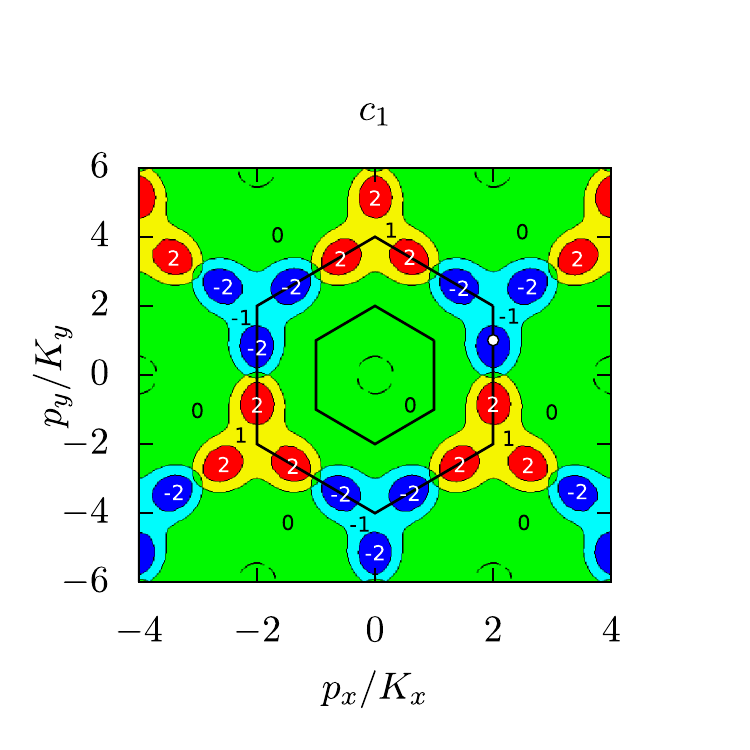}\hfill{}\includegraphics[width=0.45\textwidth]{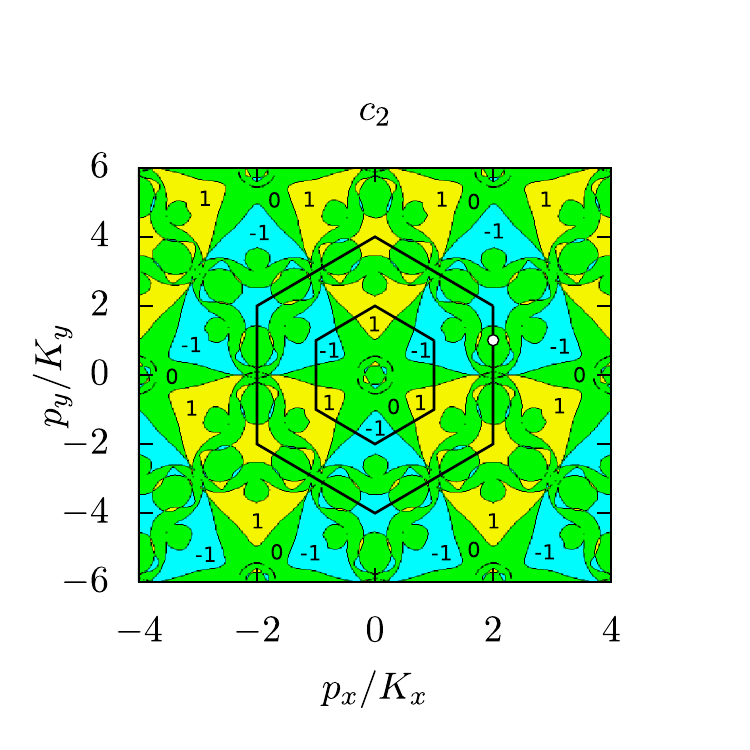}

\protect\caption{(Color online) Chern number dependence on the recoil momentum in the case $\varepsilon=0.5J$
and $J_{2}=0.5J$. \emph{Left}: Chern number $c_{1}$ of the lowest
band. \emph{Right}: Chern number $c_{2}$ of the middle band. The
color scheme and labeling are the same as in the figs.~\ref{fig:semi-metal-121}
and \ref{fig:insulator-101}. The white point is $\bm{p}=(2K_{x},K_{y})$
where the Chern numbers are $c_{1}=-2$, $c_{2}=0$ and $c_{3}$ (see
the spectrum in the Fig.~ \ref{fig:spectrum-121-202}).}

\label{fig:metal-202}
\end{figure}

There are more types of Chern phases when the coupling $J_{2}$ is
larger than in the previous discussion and comparable to the on-site
energy $\varepsilon$. For $\varepsilon=J_{2}=0.5J$ we find insulating
phases with Chern numbers $\{\pm1,\pm1,\mp2\}$ and metallic phases
with Chern numbers $\{\pm2,0,\mp2\}$ (Fig.~\ref{fig:metal-202}).
For example in the point $\bm{p}=2\bm{K}$ we get Chern numbers $c_{1}=c_{2}=-1$
and $c_{3}=2$ with band gaps $\Delta_{12}\approx0.61J$ and $\Delta_{23}\approx0.54J$.
The width of the lower two bands are around $3 J$, while the band width of the
highest band is about $1.5 J$.
Another interesting point is $\bm{p}=(2K_{x},K_{y})$ where the Chern
numbers are $c_{1}=-2$, $c_{2}=0$ and $c_{3}=2$ (white point in
the Fig.~\ref{fig:metal-202}). The bulk spectrum in this point is
given in the Fig.~\ref{fig:spectrum-121-202}. Note that there is
a gap $\Delta_{13}\approx1.35J$ between the lowest and highest bands.
In this gap there is a middle band with a zero Chern number. By setting
the Fermi energy in this gap one gets the Chern metallic phase with
the Chern number $c_{1}=-2$.

To summarize the numeric analysis for $J_2 \neq 0$,
the typical nontrivial Chern number distributions over the bands are
$\{ 0, \pm 1, \mp 1 \}$, $\{ \pm 1, \mp 1, 0 \}$, $\{ \pm 1, 0, \mp 1 \}$,
$\{ \pm 1, \pm 1, \mp 2 \}$, $\{ \pm 2, \mp 1, \mp 1 \}$ and
$\{ \pm 2, 0, \mp 2 \}$. One can also find the case $\{ \pm 1, \mp 2, \pm 1 \}$,
which is typical for $J_2 = 0$. For smaller $J_2$ compared to $J$ and $\varepsilon$,
one usually gets Chern numbers up to 1 in modulus.
For larger Chern numbers (up to 2 in modulus), one needs to make
the NNN-hopping $J_2$ be comparable to the on-site energy mismatch $\varepsilon$.

\begin{figure}
\includegraphics[width=0.45\textwidth]{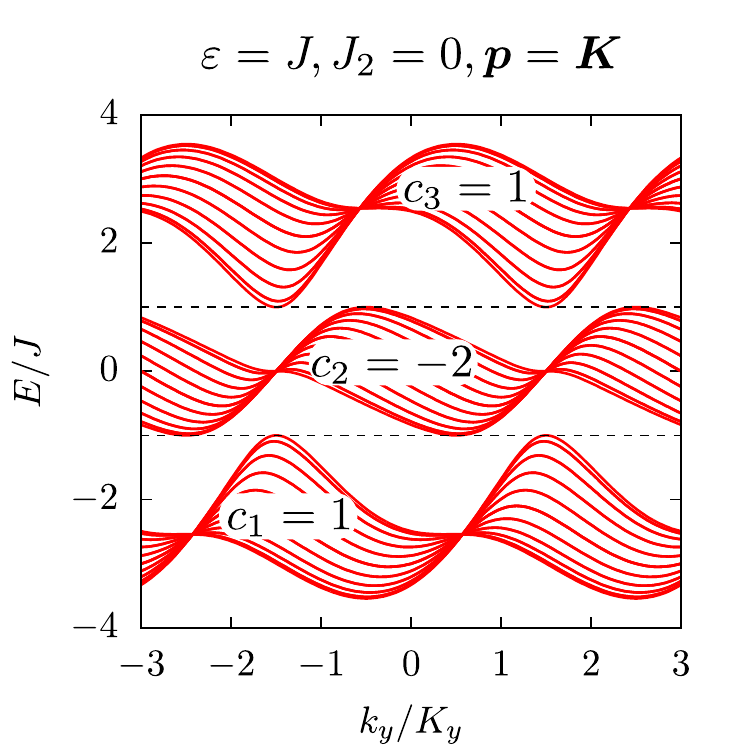}\hfill{}\includegraphics[width=0.45\textwidth]{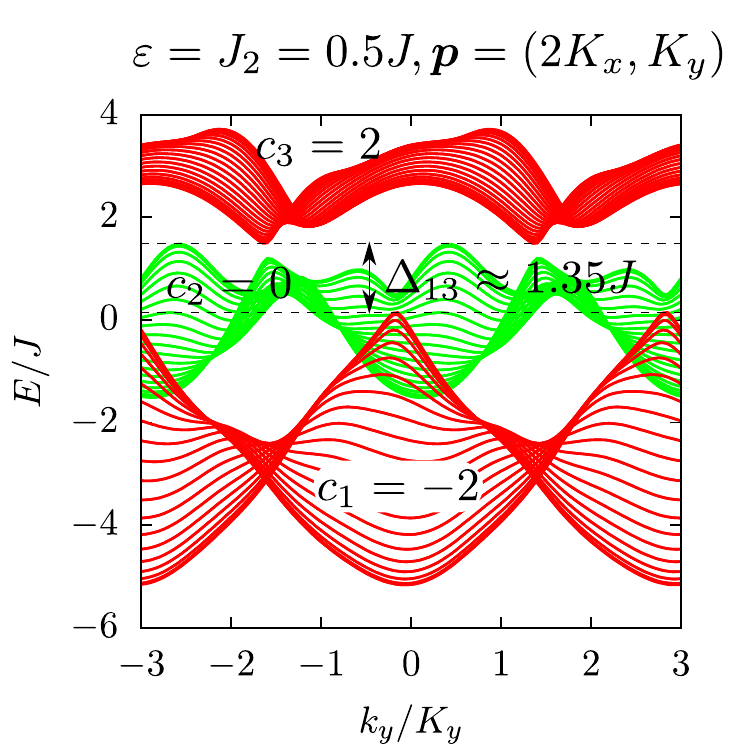}

\protect\caption{(Color online) Bulk lattice spectrum projected along $k_y$ for a number of different $k_x$ values in the range $-K_x\le k_x\le K_x$. \emph{Left}: the spectrum for the recoil momentum $\bm{p}=\bm{K}$
in the absence of the NNN-coupling ($J_2=0$) and for $\varepsilon = J$  corresponding to the parameters used in the Fig.~\ref{fig:semi-metal-121}.
In that case there is no energy gap in the spectrum, but different energy bands do not directly touch each other. 
A topological semi-metal phase is formed if the atoms fill the first energy band or the first two bands. \emph{Right}:
the spectrum for the recoil momentum $\bm{p}=(2K_{x},K_{y})$ in the case where $\varepsilon=0.5J$ and $J_{2}=0.5J$ corresponding to the phase diagram shown in
Fig.~\ref{fig:metal-202}.
Now there are two bands with non-zero Chern numbers $\pm2$ separated by a quasi-gap $\Delta_{13}\approx1.35J$
containing a middle band with a zero Chern number.}

\label{fig:spectrum-121-202}
\end{figure}

\section{Analytical Chern number calculation}

Analytic Chern number calculation is based on
integration of a Berry connection around each singularity point.
The Berry connection of the $n$-th band is defined as \cite{Goldman2014RPP,Xiao2010}
\begin{equation}
\bm{A}_{n}(\bm{k})={\rm i}\langle u_{\bm{k},n}|\nabla_{\bm{k}}|u_{\bm{k},n}\rangle,\label{eq:Berry-connection}
\end{equation}
where $|u_{\bm{k},n}\rangle$ denotes the $n$-th eigenvector of
the matrix (\ref{eq:H-matrix}). One can express the Berry curvature
(\ref{eq:Berry-curvature}) as the $z$ component of the curl $\bm{B}_{n}=\nabla\times\bm{A}_n$,
namely $F_{n}(\bm{k})=\bm{e}_{z}\cdot\bm{B}_{n}$.
Using the Stoke's
theorem we change the integral featured in Eq.(\ref{eq:Chern-number})
over the FBZ to a contour integral around the FBZ,
\[
\frac{1}{2\pi}\int_{{\rm FBZ}}{\rm d}^{2}k\, F_{n}(\bm{k})\rightarrow\frac{1}{2\pi}\oint_{{\rm FBZ}}{\rm d}\bm{k}\cdot\bm{A}_{n}-\frac{1}{2\pi}\sum\oint_{{\rm singul}}{\rm d}\bm{k}\cdot\bm{A}_{n}\,,
\]
where the last term excludes any contribution due to unphysical gauge-dependent
singular points of the Berry connection \cite{Hatsugai2005,Hafezi2008,Goldman2013}. Since
the $\bm{k}$-space Hamiltonian $H(\bm{k})$, given by Eq.~(\ref{eq:H-matrix}) or (\ref{eq:H-matrix-simplified}), and its eigenstates
are periodic in the FBZ, $\bm{A}_{n}$ is also periodic. Thus the
contour integral around the FBZ (the first term on the r.h.s. of the
above equation) is zero. Consequently the Chern number (\ref{eq:Chern-number})
can be calculated by integrating $\bm{A}_{n}$ around each excluded
singular point \cite{Goldman2013}:
\begin{equation}
c_{n}=\frac{1}{2\pi}\sum\oint_{{\rm singul}}{\rm d}\bm{k}\cdot\bm{A}_{n}\,,\label{eq:Chern-number-analytic}
\end{equation}
where the sum is over all singular points in the FBZ.

Let us summarize our analytical results, details being presented in the Appendix.
For the case where the recoil momentum coincides with the inverse lattice vector ($\bm{p}=\bm{G}$)
we always have trivial phase with all three Chern numbers equal to zero.
For the semi-metal case (Fig.\ref{fig:semi-metal-121}) with no NNN-hopping and $\bm{p}=\bm{K}$
we find two phases, depending on the mismatch $\varepsilon$ of the on-site energies.
If $\varepsilon < \varepsilon_0 = \frac{3 \sqrt{2}}{2} J$, we get Chern semi-metal phase with
Chern numbers $\{ 1, -2, 1 \}$. If $\varepsilon > \varepsilon_0$, we get a trivial phase $\{ 0,0,0 \}$.
In this way at larger mismatch between the on-site energies the topological
phenomena disappear. This is in agreement with the numerical calculation presented in the previous Section.

It is possible to apply this method for other values of the recoil momenta $\bm{p}$ and for a general non-symmetric
case with the NNN-hoppings. In such calculations one needs to diagonalize
the matrices of the size at most $2\times2$.  Yet generally ordering of the eigenvalues
might be a quite involved task, especially if they depend
on more than one parameter.

\section{Concluding remarks}

In conclusion, we have considered a two-dimensional dice lattice operating
in a tight-binding regime. The laser-assisted nearest neighbor transitions 
are accompanied by the momentum recoils. This allows one to engineer
staggered synthetic magnetic fluxes and thus facilitates
realization of topologically nontrivial band structures. Real
valued next nearest neighbor transitions -- although not necessary in principle
to reach the topological regime -- may also be present and contribute
to the richness of the obtained topological phases. The considered
dice lattice represents a triangular Bravais lattice with a three-site
basis consisting of a hub site connected to two rim sites, providing
three energy bands. Thus our model
can be interpreted as a generalization of the paradigmatic Haldane
model which is reproduced if one of the two rim sub-lattices is eliminated.
We have demonstrated that the proposed upgrade of the Haldane model
creates a significant added value such as (i) an easy access to topological
semimetal phases relying on only the nearest neighbor coupling and (ii)
enhanced topological band structures featuring Chern numbers higher
than one and thus providing access to physics beyond the usual quantum 
Hall effect. The numerical analysis have been supported by an analytical
scheme based on the study of singularities in the Berry connection.

\appendix

\section{Details on analytical Chern number calculation}

\subsection{Momentum space Hamiltonian and its eigenstates}

Let us establish a general structure of the eigenstates for the
matrix Hamiltonian $\mathcal{H}(\bm{k})$, Eq.(\ref{eq:H-matrix-simplified}).
For this we introduce a basis of our three-level system $|s\rangle$,
with $s=0,\pm1$, and rewrite the matrix Hamiltonian in the state-vector
notation as
\begin{equation}
\mathcal{H}(\bm{k})=\sum_{s=0,\pm1}|s\rangle d_{s}(\bm{k})\langle s|+\sum_{s=\pm1}\left(|s\rangle g_{s}(\bm{k}){\rm e}^{{\rm i}s\alpha_{s}(\bm{k})}\langle0|+{\rm H.\, c.}\right),\label{eq:Hamiltonian-in-k-point}
\end{equation}
where $d_{s}(\bm{k})$ stands for the diagonal matrix elements: 
\begin{equation}
d_{s}(\bm{k})=s\varepsilon+2J_{2}f(\bm{k}-s\bm{p})\,.\label{eq:d_s}
\end{equation}
The off-diagonal matrix elements 
\begin{equation}
Jg(\bm{k}\mp\bm{p}/2)=g_{\pm}(\bm{k}){\rm e}^{{\rm i}\alpha_{\pm}(\bm{k})}\label{eq:g_pm}
\end{equation}
have been represented in terms of their amplitudes $g_{\pm1}(\bm{k})\equiv g_{\pm}(\bm{k})$
and phases $\alpha_{\pm1}(\bm{k})\equiv\alpha_{\pm}(\bm{k})$. 

Since there is no coupling between the A and C sub-lattices, one can
perform a $\bm{k}$-dependent unitary transformation eliminating the
phase factors
\[
|s\rangle\to|s,\bm{k}\rangle=|s\rangle{\rm e}^{{\rm i}s\alpha_{s}(\bm{k})},\qquad s=\pm1,
\]
and leave the basis vector $|0\rangle$ unchanged ($|0\rangle=|0,\bm{k}\rangle$).
In the new basis the Hamiltonian (\ref{eq:Hamiltonian-in-k-point})
is characterized by real and symmetric matrix elements. Its eigenvectors
can be cast in terms of these vectors with real coefficients $C_{n,s}(\bm{k})$:
\begin{equation}
|u_{\bm{k},n}\rangle=\sum_{s=0,\pm1}C_{n,s}(\bm{k})|s,\bm{k}\rangle\equiv\sum_{s=0,\pm1}|s\rangle C_{n,s}(\bm{k}){\rm e}^{{\rm i}s\alpha_{s}(\bm{k})}\,,\label{eq:analytic-eigenvector}
\end{equation}
Combining Eqs.~(\ref{eq:Berry-connection}) and (\ref{eq:analytic-eigenvector}),
one arrives at the following expression for the Berry connection
\begin{equation}
\bm{A}_{n}(\bm{k})=-\sum_{s=\pm1}sC_{n,s}^{2}(\bm{k})\nabla_{\bm{k}}\alpha_{s}(\bm{k}).\label{eq:Berry-connection-analytic}
\end{equation}
This result together with Eq.~(\ref{eq:Chern-number-analytic}) will
be subsequently used in finding the Chern numbers.

\subsection{Determination of the Chern numbers: General}

To determine the Chern number given by (\ref{eq:Chern-number-analytic}),
one needs to study a behavior of the vector potential at its singular
points. Singularities of the vector potential can emerge at the points
where the phase of the coupling matrix element $g_{\pm}(\bm{k}){\rm e}^{{\rm i}\alpha_{\pm}(\bm{k})}$
given by Eq.~(\ref{eq:g_pm}) is undefined. This happens if
the function $g(\bm{k}-\bm{p}_{\pm}/2)$ goes to zero. The function
$g(\bm{k})$ given by Eq.~(\ref{eq:f-and-g-functions}) is zero at
the corners of the FBZ, namely at two inequivalent points $\bm{K}$
and $\bm{K}^{\prime}$. Thus there are two pairs of points 
\begin{equation}
\bm{K}_{\pm}=\pm\bm{p}/2+\bm{K}\,,\quad\bm{K}_{\pm}^{\prime}=\pm\bm{p}/2+\bm{K}^{\prime}\,.\label{eq:K_pm}
\end{equation}
at which the function $g(\bm{k}\mp\bm{p}/2)$ goes to zero and its
phase $\alpha_{\pm}(\bm{k})$ is undefined. Let us determine the coupling
matrix element $g_{\pm}(\bm{k}){\rm e}^{{\rm i}\alpha_{\pm}(\bm{k})}$
in a vicinity of these points. Combining Eqs.~(\ref{eq:f-and-g-functions})
and (\ref{eq:g_pm}), the amplitude and phase of the coupling element
reads up to the first-order in the displacement vector $\bm{q}$, i.e.
for $qa\ll1$ with $q=|\bm{q}|$: 
\begin{eqnarray}
g_{\pm}(\bm{K}_{\pm}+\bm{q}) & \approx\frac{3}{2}qaJ\,,\quad\alpha_{\pm}(\bm{K}_{\pm}+\bm{q})\approx\frac{\pi}{3}-\varphi,\label{eq:g-expansion-K}\\
g_{\pm}(\bm{K}_{\pm}^{\prime}+\bm{q}) & \approx\frac{3}{2}qaJ\,,\quad\alpha_{\pm}(\bm{K}_{\pm}^{\prime}+\bm{q})\approx-\frac{\pi}{3}+\varphi,\label{eq:g-expansion-Kprime}
\end{eqnarray}
where $\varphi$ is a phase of the complex number $q_{x}+{\rm i}q_{y}=q{\rm e}^{{\rm i}\varphi}$.
Integrating over a small circle centered at $\mathbf{q}=0$ surrounding
each singular point of the phase, one finds:
\[
\oint_{\mathbf{\left|q\right|}\rightarrow0}{\rm d}\bm{q}\cdot\nabla_{q}\alpha_{\pm}(\bm{K}_{\pm}+\bm{q})=-2\pi\,,
\]
\[
\oint_{{\rm \mathbf{\left|q\right|}\rightarrow0}}{\rm d}\bm{q}\cdot\nabla_{q}\alpha_{\pm}(\bm{K}_{\pm}^{\prime}+\bm{q})=2\pi\,,
\]
where the signs are different due to the opposite phases
in Eqs.~(\ref{eq:g-expansion-K}) and (\ref{eq:g-expansion-Kprime}).
These equations together with Eqs.~(\ref{eq:Chern-number-analytic})
and (\ref{eq:Berry-connection-analytic}) provide the following result
for the Chern number
\begin{equation}
c_{n}=\sum_{s=\pm1}s\left[C_{n,s}^{2}(\bm{K}_{s})-C_{n,s}^{2}(\bm{K}_{s}^{\prime})\right]\,,\label{eq:Chern-number-analytic-1}
\end{equation}
with $\bm{K}_{\pm1}\equiv\bm{K}_{\pm}$ and $\bm{K}_{\pm1}^{\prime}\equiv\bm{K}_{\pm}^{\prime}$.
Therefore to find the Chern number one needs to determine the coefficients
$C_{n,s}$ entering the state-vector at the points $\bm{K}_{\pm}$
and $\bm{K}_{\pm}^{\prime}$. If $C_{n,\pm}^{2}=1$, the corresponding
singular point contributes to the Chern number of the $n$-th band.
In the following we shall analyze two different situations.

\subsection{Determination of the Chern numbers: Specific cases}

Since the Hamiltonian $\mathcal{H}(\bm{k})$ given by Eq.~(\ref{eq:H-matrix-simplified}) or (\ref{eq:Hamiltonian-in-k-point}) has a symmetry
$(\varepsilon\to-\varepsilon,\,\mathcal{H}\to\mathcal{H})$, we consider
only the case where $\varepsilon>0$.

\subsubsection{The case where $\bm{p}=\bm{G}$}

Suppose first that the difference in the recoil momenta coincides
with the inverse lattice vector $\bm{p}=\bm{G}$. In that case the
coupling completely vanishes both for $\bm{k}=\bm{K}_{\pm}$ and also
for $\bm{k}=\bm{K}_{\pm}^{\prime}$. At these points $g(\bm{k}-\bm{p}/2)=g(\bm{k}+\bm{p}/2)=0$,
so all the states $|s\rangle$ ($s=0,\pm1$) are decoupled, and thus
the eigenstates are the bare states $|s\rangle$. The corresponding
eigen-energies of the matrix Hamiltonian $\mathcal{H}(\bm{k})$, Eq.(\ref{eq:Hamiltonian-in-k-point}),
coincide with its diagonal elements $d_{s}(\bm{k})$ for $\bm{k}=\bm{K}_{\pm}$
and $\bm{k}=\bm{K}_{\pm}^{\prime}$. Since $\bm{p}=\bm{G}$,
one has $f(\bm{k}-\bm{p})=f(\bm{k})=f(\bm{k}+\bm{p})$, giving $d_{s}(\bm{k})=s\varepsilon+2J_{2}f(\bm{k})$.
Therefore the eigenstates are ordered in the same manner $d_{+1}(\bm{k})>d_{0}(\bm{k})>d_{-1}(\bm{k})$ 
both for $\bm{k}=\bm{K}_{\pm}$ and also $\bm{k}=\bm{K}_{\pm}^{\prime}$, giving
$C_{n,s}^{2}(\bm{K}_{s})=C_{n,s}^{2}(\bm{K}_{s}^{\prime})$ with $s=\pm1$.
As a result, the Chern number given by Eq.~(\ref{eq:Chern-number-analytic-1})
is identically equal to zero, and the
system does not exhibit any topologically non-trivial phases. This
is because for $\bm{p}=\bm{G}$ the flux over the rhombic plaquettes $\Phi_{i}=\pm\bm{p}\cdot\bm{a}_{i}$
is zero (modulo $2\pi$), and there is no breaking of the time-reversal
symmetry.

\subsubsection{The case where $\bm{p}=\bm{K}$}

\begin{figure}
\begin{centering}
\includegraphics[width=0.40\textwidth]{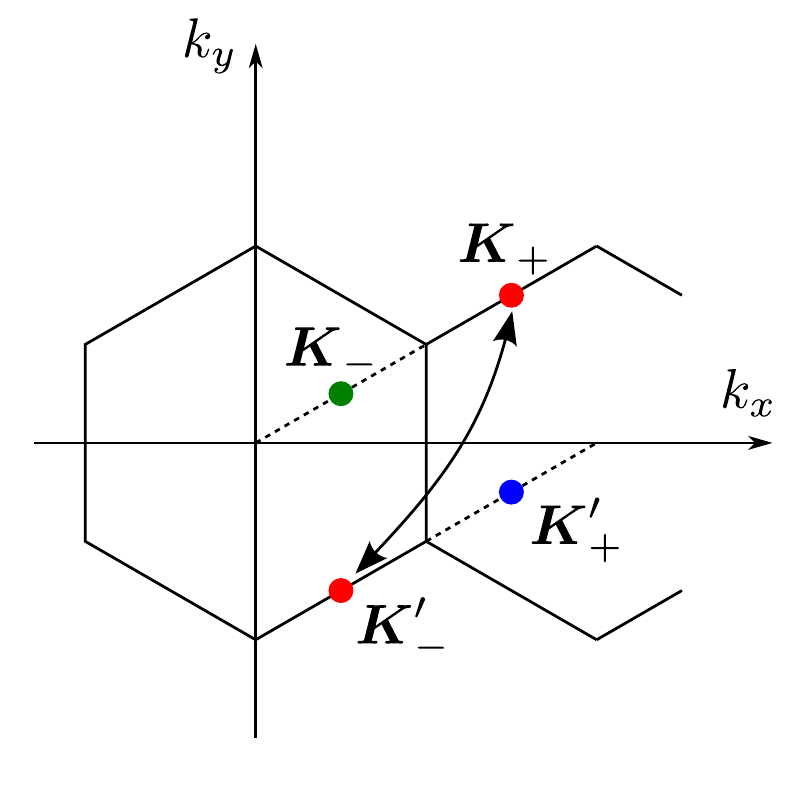}
\par\end{centering}

\protect\caption{(Color online) The phase singularity points $\bm{K}_{\pm}$ and $\bm{K}_{\pm}^{\prime}$ of the 
coupling matrix elements $g(\bm{k}\mp\bm{p}/2)$ for $\bm{p}=\bm{K}$.
The points $\bm{K}_{+}$ and $\bm{K}_{-}^{\prime}$ are equivalent. They are shown by red
dots connected with a double arrow.}

\label{fig:singul}
\end{figure}
As another illustration we pick the recoil momentum $\bm{p}=\bm{K}$ and take  $J_{2}=0$. In that
case the Chern numbers have been numerically found to be $c_{1}=1$, $c_{2}=-2$
and $c_{3}=1$, see Fig.~\ref{fig:semi-metal-121}. By taking $\bm{p}=\bm{K}$ the phase singularities of the
coupling elements $g(\bm{k}\mp\bm{p}/2)$ emerge at the points $\bm{K}_{\pm}=\pm\bm{K}/2+\bm{K}$
and $\bm{K}_{\pm}^{\prime}=\pm\bm{K}/2+\bm{K}^{\prime}$, as one can see in Fig.~\ref{fig:singul}.
Furthermore, the point $\bm{k}=\bm{K}_{+}$ is equivalent to the point
$\bm{k}=\bm{K}_{-}^{\prime}$. For the latter two points we have $g(\bm{k}-\bm{p}/2)=g(\bm{k}+\bm{p}/2)=0$,
so there are no coupling matrix elements. Since $J_{2}=0$, the Hamiltonian (\ref{eq:Hamiltonian-in-k-point})
at these points is simply 
\begin{equation}
\mathcal{H}(\bm{K}_{+})=\mathcal{H}(\bm{K}_{-}^{\prime})=\varepsilon\sum_{s=\pm1}s|s\rangle\langle s|\,,\label{eq:H-in-Kplus}
\end{equation}
so the diagonal energies entering the Hamiltonian
(\ref{eq:Hamiltonian-in-k-point}) are $d_{s}(\bm{k})=s\varepsilon$. 

Eigenvalues, ordered from the lowest to the highest, are $E_{1}(\bm{K}_{+})=E_{1}(\bm{K}_{-}^{\prime})=-\varepsilon$,
$E_{2}(\bm{K}_{+})=E_{2}(\bm{K}_{-}^{\prime})=0$ and $E_{3}(\bm{K}_{+})=E_{3}(\bm{K}_{-}^{\prime})=\varepsilon$.
There is no degeneracy for $\varepsilon>0$ and the coefficients
$C_{n,+}(\bm{K}_{+})$ and $C_{n,-}(\bm{K}_{-}^{\prime})$ do not
change if one increases $\varepsilon$. The only non-zero coefficients
contributing to the Chern numbers read
\begin{equation}
C_{3,+}(\bm{K}_{+})=C_{1,-}(\bm{K}_{-}^{\prime})=1.\label{eq:non-zero-coeffs-1}
\end{equation}

\begin{figure}
	\includegraphics[width=0.45\textwidth]{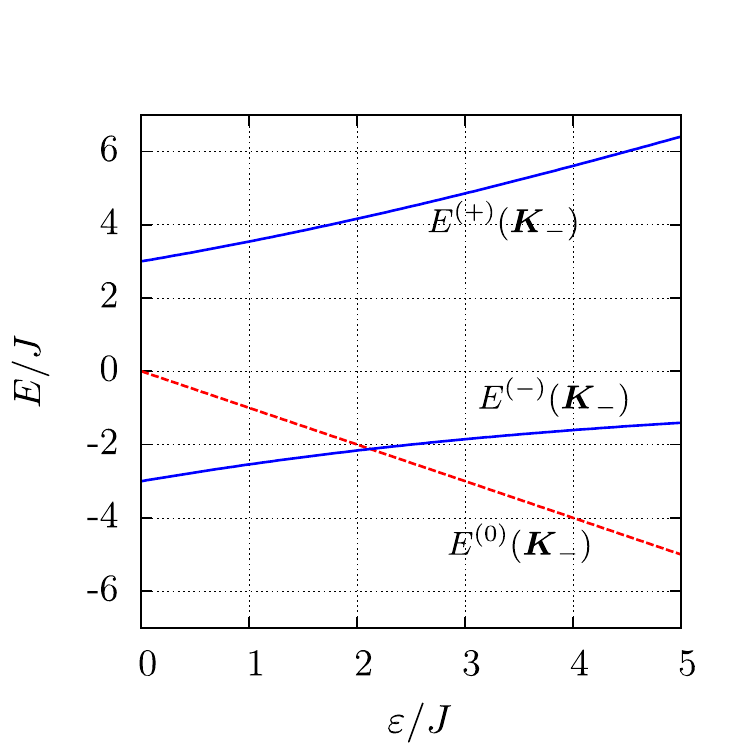}\hfill{}\includegraphics[width=0.45\textwidth]{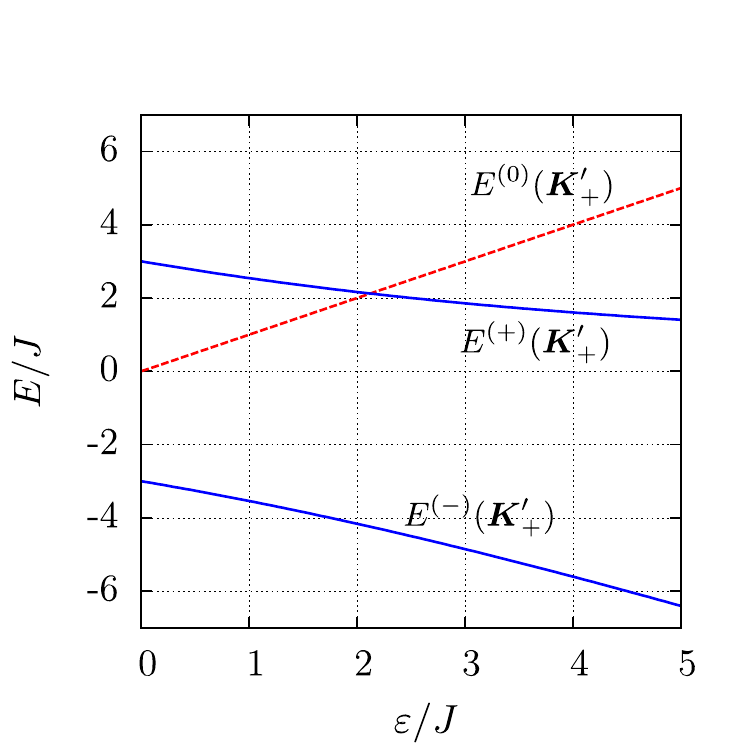}
	
	\protect\caption{Dependence of eigenvalues of the Hamiltonian $H(\bm{k})$ on the on-site
		energy $\varepsilon$ for $\bm{p}=\bm{K}$ in the absence of the next-nearest neighbour coupling.
		The eigenvalue $E^{(0)}$ is plotted
		in red dashed lines to distinguish it from the other eigenvalues $E^{(\pm)}$
		\emph{Left}: eigenvalues
		at the point $\bm{k}=\bm{K}_{-}$. \emph{Right}: eigenvalues at the
		point $\bm{k}=\bm{K}_{+}^{\prime}$. The eigenvalue crossing point $\varepsilon=\frac{3\sqrt{2}}{2}J\equiv\varepsilon_{0}$ corresponds to a transition from
		a topological semimetal phase on the left to a trivial phase on the right.
	}

	\label{fig:evals}
\end{figure}
For the point $\bm{k}=\bm{K}_{-}$ the non-diagonal matrix elements
of (\ref{eq:Hamiltonian-in-k-point}) are $Jg(\bm{k}+\bm{p}/2)=0$
and $Jg(\bm{k}-\bm{p}/2)=3J$. Similarly for the point $\bm{k}=\bm{K}_{+}^{\prime}$
these elements are $Jg(\bm{k}-\bm{p}/2)=0$ and $Jg(\bm{k}+\bm{p}/2)=3J$.
Thus the Hamiltonian (\ref{eq:Hamiltonian-in-k-point}) at these points
is
\begin{eqnarray}
H(\bm{K}_{-}) & =\varepsilon\sum_{s=\pm1}s|s\rangle\langle s|+3J\big(|0\rangle\langle+|+|+\rangle\langle0|\big),\label{eq:H-in-Kminus}\\
H(\bm{K}_{+}^{\prime}) & =\varepsilon\sum_{s=\pm1}s|s\rangle\langle s|+3J\big(|0\rangle\langle-|+|-\rangle\langle0|\big).\label{eq:H-in-Kplusprime}
\end{eqnarray}
Eigenvalues of the $H(\bm{K}_{-})$
are $E^{(0)}(\bm{K}_{-})=-\varepsilon$ and $E^{(\pm)}(\bm{K}_{-})=\frac{1}{2}\left(\varepsilon\pm\sqrt{\varepsilon^{2}+36J^{2}}\right)$, and whose of $H(\bm{K}_{+}^{\prime})$ are 
$E^{(0)}(\bm{K}_{+}^{\prime})=\varepsilon$ and $E^{(\pm)}(\bm{K}_{+}^{\prime})=\frac{1}{2}\left(-\varepsilon\pm\sqrt{\varepsilon^{2}+36J^{2}}\right)$.
They are plotted in Fig.~\ref{fig:evals}. For $\varepsilon=\frac{3\sqrt{2}}{2}J\equiv\varepsilon_{0}$
there are degeneracies $E^{(0)}(\bm{K}_{-})=E^{(-)}(\bm{K}_{-})=-\varepsilon_{0}$
and $E^{(0)}(\bm{K}_{+}^{\prime})=E^{(+)}(\bm{K}_{+}^{\prime})=\varepsilon_{0}$.
The eigenvalues change their order at the crossing point $\varepsilon=\varepsilon_{0}$, as one can see in
Fig.~\ref{fig:evals}.

Let us first consider the case $0<\varepsilon<\varepsilon_{0}$. The eigenvalues
of $H(\bm{K}_{-})$ are in the increasing order: $E_{1}(\bm{K}_{-})=E^{(-)}(\bm{K}_{-})$,
$E_{2}(\bm{K}_{-})=E^{(0)}(\bm{K}_{-})$ and $E_{3}(\bm{K}_{-})=E^{(+)}(\bm{K}_{-})$.
On the other hand, coefficients required for the Chern number calculation
are $C_{1,-}(\bm{K}_{-})=0$, $C_{2,-}(\bm{K}_{-})=1$ and $C_{3,-}(\bm{K}_{-})=0$.
Similarly $H(\bm{K}_{+}^{\prime})$ gives the
eigenvalues $E_{1}(\bm{K}_{+}^{\prime})=E^{(-)}(\bm{K}_{+}^{\prime})$,
$E_{2}(\bm{K}_{+}^{\prime})=E^{(0)}(\bm{K}_{+}^{\prime})$ and $E_{3}(\bm{K}_{+}^{\prime})=E^{(+)}(\bm{K}_{+}^{\prime})$
and the coefficients $C_{1,+}(\bm{K}_{+}^{\prime})=0$, $C_{2,+}(\bm{K}_{+}^{\prime})=1$
and $C_{3,+}(\bm{K}_{+}^{\prime})=0$. Combining this result together
with (\ref{eq:non-zero-coeffs-1}) we collect four non-zero coefficients:
$C_{3,+}(\bm{K}_{+})$, $C_{1,-}(\bm{K}_{-}^{\prime})$, $C_{2,-}(\bm{K}_{-})$
and $C_{2,+}(\bm{K}_{+}^{\prime})$. Substituting them into 
Eq.~(\ref{eq:Chern-number-analytic-1}), we get the Chern numbers for each energy band
\begin{eqnarray}
c_{1} & =C_{1,-}^{2}(\bm{K}_{-}^{\prime})=1,\label{eq:analytic-c-1}\\
c_{2} & =-C_{2,+}^{2}(\bm{K}_{+}^{\prime})-C_{2,-}^{2}(\bm{K}_{-})=-2,\label{eq:analytic-c-2}\\
c_{3} & =C_{3,+}^{2}(\bm{K}_{+})=1.\label{eq:analytic-c-3}
\end{eqnarray}
This result agrees with the numerical analysis presented in Fig.~\ref{fig:semi-metal-121}.

Now let us consider the case $\varepsilon>\varepsilon_{0}$. From
the Fig.~\ref{fig:evals} we see that the eigenvalues are reordered
as $E_{1}(\bm{K}_{-})\rightarrow E_{2}(\bm{K}_{-})$, $E_{2}(\bm{K}_{+}^{\prime})\rightarrow E_{3}(\bm{K}_{+}^{\prime})$,
so non-zero coefficients are $C_{3,+}(\bm{K}_{+})$,
$C_{1,-}(\bm{K}_{-}^{\prime})$, $C_{1,-}(\bm{K}_{-})$ and $C_{3,+}(\bm{K}_{+}^{\prime})$.
Using Eq.~(\ref{eq:Chern-number-analytic-1}), one can see that the Chern numbers of all bands are now zero:
\begin{eqnarray}
c_{1} & =-C_{1,-}^{2}(\bm{K}_{-})+C_{1,-}^{2}(\bm{K}_{-}^{\prime})=0,\label{eq:analytic-c-1-zero}\\
c_{2} & =0,\label{eq:analytic-c-2-zero}\\
c_{3} & =C_{3,+}^{2}(\bm{K}_{+})-C_{3,+}^{2}(\bm{K}_{+}^{\prime})=0.\label{eq:analytic-c-3-zero}
\end{eqnarray}
Thus there is a topological phase transition at $\varepsilon=\frac{3\sqrt{2}}{2}J$ 
corresponding to the eigenvalue crossing in Fig.~\ref{fig:evals}.

\begin{acknowledgments}
This research was supported by the Research Council of Lithuania (Grant No. MIP-082/2012). 
IBS was partially supported by the ARO's atomtronics MURI, the AFOSR's Quantum Matter MURI, NIST, 
and the NSF through the PFC at the JQI.
\end{acknowledgments}

\bibliography{main}

\end{document}